\documentclass[12pt]{elsarticle}

\usepackage{mathrsfs}
\usepackage{amsmath}
\usepackage{amssymb}

\usepackage{graphicx}
\usepackage{multirow}

\usepackage{doi}
\usepackage{hyperref}
\usepackage[numbers]{natbib}

\usepackage{xcolor}

\biboptions{sort&compress}

\allowdisplaybreaks

\newtheorem{remark}{Remark}

\journal{AIP}

\begin{document}
\begin{frontmatter}

\title{ Symmetries and conservation laws of the one-dimensional shallow
water magnetohydrodynamics equations in~Lagrangian coordinates}

\author[sutaddress]{S.~V. Meleshko\corref{mycorrespondingauthor}}

\cortext[mycorrespondingauthor]{Corresponding author}

\ead{sergey@math.sut.ac.th}

\author[kiamaddress]{V.~A. Dorodnitsyn}

\ead{Dorodnitsyn@keldysh.ru,dorod2007@gmail.com}

\author[sutaddress]{E.~I. Kaptsov}

\ead{evgkaptsov@math.sut.ac.th}

\address[sutaddress]{School of Mathematics, Institute of Science, \\
 Suranaree University of Technology, 30000, Thailand}

\address[kiamaddress]{Keldysh Institute of Applied Mathematics,\\
 Russian Academy of Science, Miusskaya Pl.~4, Moscow, 125047, Russia}
\begin{abstract}
Symmetries of the one-dimensional shallow water magnetohydrodynamics
equations~(SMHD) in Gilman's approximation are studied.
The~SMHD equations are considered in case of a plane and uneven bottom
topography in Lagrangian and Eulerian coordinates. Symmetry classification
separates out all bottom topographies which yields substantially different admitted
symmetries.

The~SMHD equations in Lagrangian coordinates were reduced to a single
second order~PDE. The Lagrangian formalism and~Noether's theorem
are used to construct conservation laws of the~SMHD equations. Some
new conservation laws for various bottom topographies are obtained.
The results are also represented in Eulerian coordinates.

Invariant and partially invariant solutions are constructed.
\end{abstract}
\begin{keyword}
shallow water \sep magnetohydrodynamics \sep Lagrangian coordinates
\sep Lie symmetries \sep Conservation laws \sep exact invariant
solutions
\end{keyword}
\end{frontmatter}

\section{Introduction}

The Gilman shallow water magnetohydrodynamics model (SMHD)~\cite{bk:SWMHD_Gilman2000}
is applied for describing the global dynamics of the solar tachocline, a thin layer
at the base of the solar convection zone. It was demonstrated that
the tachocline can be regarded as two-dimensional shallow water in
the presence of a flat magnetic field~\cite{bk:SWMHD_Gilman2000}.
This result significantly increased the utility of the shallow
water approximation for astrophysical problems
and led to numerous publications on~SMHD.

Properties of the SMHD equations as a nonlinear system of hyperbolic
conservation laws were studied in~\cite{bk:SWMHD_DeSterck2001}, where
the foundations laid for constructing accurate shock-capturing
numerical schemes for the~SMHD equations.
In~\cite{bk:SWMHD_Dellar2003}, a~Hamiltonian formulation of the~SMHD was used,
and a dispersive system, based on the model~\cite{bk:SWMHD_Gilman2000} was constructed
to avoid unphysical cusp-like singularities in finite amplitude magnetogravity
waves. In~\cite{bk:SWMHD_Zeitlin2013}, it was shown that the~SMHD
model may be systematically derived by vertical averaging of the full~MHD
equations for the rotating magnetofluid under the gravity influence.
A number of properties of the model~\cite{bk:SWMHD_Gilman2000}
was investigated in~\cite{bk:SWMHD_Karelsky2014}, where self-similar
discontinuous and continuous solutions were found. The local well-posedness
in time of the one-dimensional model was proven in~\cite{bk:SWMHD_SHIUE2013215}.
The structural stability of shock waves and current-vortex sheets
in the~SMHD was studied in~\cite{bk:SWMHD_Trakhinin2020}.

Based on the model~\cite{bk:SWMHD_Gilman2000}, a large number of
numerical discretizations have been constructed~\cite{bk:SWMHD_Touma2010,bk:SWMHD_Winters2016,bk:SWMHD_AhmedZia2019,bk:SWMHD_JunmingHuazhong2021},
including discretizations for the case of non-flat bottom topography.

\smallskip{}

Although the model~\cite{bk:SWMHD_Gilman2000} is the very common,
it is not the only~SMHD model. While the most works
on~SMHD consider initially toroidal magnetic fields, in neutron star-related
applications, however, it makes sense to consider initially vertical
magnetic fields. The~SMHD equations in the external vertical magnetic
field were covered in~\cite{bk:SWMHD_Heng2009,bk:SWMHD_PetrosyanKlimachkov2016,bk:SWMHD_KlimachkovPetrosyan2017,bk:SWMHD_KlimachkovPetrosyan2017b,bk:SWMHD_Hindle2019,bk:SWMHD_Petrosyan2020}.
The extended~SMHD model in the presence of vorticity and magnetic
currents was proposed in~\cite{bk:SWMHD_AlonsoOran2021}. Motion
of an ideal fluid flow under the influence of a constant external
field~(gravitational or magnetic) may also be modeled by means of
modified shallow water equations~\cite{bk:KarelskyPetrosyan2008}
for which the authors have recently constructed conservative symmetry-preserving
finite-difference schemes~\cite{bk:KaptDorMel_ModSW2022}.

\medskip{}

Despite the many publications on the~SMHD
the symmetries of the~SMHD equations have not yet been
studied. The symmetries of the equations express important
physical properties of the model and they are closely related
with its conservation laws~\cite{bk:Ovsyannikov[1962],bk:Olver}.
They also allow finding exact solutions reducing partial differential equations~(PDEs)
to ordinary differential equations.

The most of cited research papers are devoted to the~SMHD equations in~Eulerian coordinates.
An alternative to the~Eulerian description is the~Lagrangian one, which has some advantages.
Lagrangian approach for studying motions of continuous medium is essentially based on
a description of the history of the motion of each specific particle
of the medium and it is always implied in the formulation of physical
laws~\cite{bk:Sedov[mss],bk:Batchelor2000}. It is also worth noting
that in numerical simulation of plasma physics and astrophysics it
is often easier to set boundary conditions in mass Lagrangian coordinates~\cite{bk:SamarskyPopov[1970]}.
For some models it is possible to construct a Lagrangian for the
equations~\cite{bk:GavrilyukTeshukov2001,bk:SiriwatKaewmaneeMeleshko2016,bk:KaptsovMeleshko_1D_classf[2018],DORODNITSYN2019201,DKKMM2021}~(see
also references therein), whereas there are no~Lagrangians in~Eulerian description.
Knowing the Lagrangian and the symmetries
of the equation makes it possible to derive conservation laws by a
simple algorithmic procedure~\cite{bk:Noether1918}.

\smallskip

In the present paper the authors fill this gap in the study of
symmetries of the one-dimensional~SMHD equations~\cite{bk:SWMHD_Gilman2000}.
The~SMHD equations are considered in case of uneven bottom topography
in both Lagrangian and~Eulerian coordinates.
The presence of an arbitrary bottom needs a group classification, which consists of identifying bottom topographies for which the admitted Lie algebras are extended.
The~Lagrangian formalism and~Noether's theorem allow
constructing conservation laws of the equations, including new
conservation laws for various bottom topographies.

\bigskip{}

The paper is organized as follows. Gilman's model in~Eulerian
coordinates, as well as its one-dimensional version with uneven bottom topography,
are given in~Section~\ref{sec:01}. In Section~\ref{sec:02} the model is
presented in mass Lagrangian coordinates, and it is shown that the
system reduces to a single second-order~PDE, while the remaining
equations are integrated. Complete group classification of this
equation with respect to the bottom topography is performed
in~Section~\ref{sec:03}. In Section~\ref{sec:04} the~Lagrangian
and~Hamiltonian for the~SMHD equations are found. Using~Noether's
theorem, conservation laws are obtained in mass~Lagrangian
coordinates, and their counterparts in~Eulerian coordinates are also presented.
Section~\ref{sec:05} is devoted to exact solutions. Invariant and
partially invariant solutions are obtained there. The results are
summarized in~Conclusion.

\section{The SMHD equations in Eulerian coordinates}
\label{sec:01}

In general, the SMHD model proposed in~\cite{bk:SWMHD_Gilman2000}, has the form
\begin{subequations}
\label{GilmanSys}
\begin{equation}
h_{t}+\nabla^{\prime}\cdot(h\mathbf{u})=0,\label{Gilman_ContinuityEq}
\end{equation}
\begin{equation}
\mathbf{u}_{t}=-\nabla\left(\frac{\mathbf{u}\cdot\mathbf{u}}{2}\right)+\nabla\left(\frac{\mathbf{H}\cdot\mathbf{H}}{2}\right)-(\hat{\mathbf{k}}\times\mathbf{u})\hat{\mathbf{k}}\cdot\nabla\times\mathbf{u}+(\hat{\mathbf{k}}\times\mathbf{H})\hat{\mathbf{k}}\cdot\nabla\times\mathbf{H}-g\nabla h=0,\label{Gilman_VEvol}
\end{equation}
\begin{equation}
\mathbf{H}_{t}=\nabla\times(\mathbf{u}\times\mathbf{H})+(\nabla^{\prime}\cdot\mathbf{u})\mathbf{H}-(\nabla^{\prime}\cdot\mathbf{H})\mathbf{u},\label{Gilman_HEvol}
\end{equation}
\begin{equation}
\nabla^{\prime}\cdot(h\mathbf{H})=0,\label{Gilman_DivH}
\end{equation}
\end{subequations}
 where $\mathbf{u}=(u,v,0)$ is the two-dimensional velocity vector,
$\mathbf{H}=(H^{x},H^{y},0)$ is the two-dimensional magnetic filed
vector, $\hat{\mathbf{k}}=(0,0,1)$ is the unit vector in the vertical
direction, $h$ characterizes a deviation of the free surface from
the undisturbed level, $\nabla^{\prime}\cdot$ is the horizontal divergence
operator, $\hat{\mathbf{k}}\cdot\nabla\times$ is the vertical component
of the curl operator, and the constant $g\neq0$ characterizes the
gravitational acceleration.

\medskip


Equation~(\ref{Gilman_ContinuityEq}) means that the free surface
is a material surface and particles initially on this surface remain
there. Equation~(\ref{Gilman_VEvol}) describes the evolution of
the horizontal velocity field, where it is taken into account that
the usual hydrostatic equation in the presence of a magnetic field
must be replaced by the condition
\begin{equation}
\nabla p=g\nabla h-\nabla\left(\frac{\mathbf{H}\cdot\mathbf{H}}{2}\right).
\end{equation}
Equation~(\ref{Gilman_HEvol}) describes the evolution
of the horizontal magnetic field. In contrast to the standard~MGD
equations, there is horizontal divergence of both velocity and magnetic
field allowed for, both of which arise in this system as a result
of the deformation of the free surface. Equation~(\ref{Gilman_DivH})
means that the magnetic field initially in the surface remains there
and remains locally parallel to the free surface~(a modified form
of the divergence-free condition for magnetic fields).

\begin{remark}
For convenience, the reduced magnetic
vector field~$\mathbf{H}$ instead of the magnetic vector field~$\widetilde{\mathbf{H}}$
is used, which are related as
\begin{equation}
\mathbf{H}=\frac{\widetilde{\mathbf{H}}}{\sqrt{4\pi\rho}},
\end{equation}
where $\rho$ is the density of the fluid,
assumed constant for shallow water.
\end{remark}

By analogy with the standard shallow water equations, the function~$b(x,y)$ characterizing topography of the bottom can be introduced~\cite{bk:Stoker1948}.
In coordinate form system~(\ref{GilmanSys}) with uneven bottom becomes
\begin{subequations}
\begin{equation}
h_{t}+uh_{x}+hu_{x}+vh_{y}+hv_{y}=0,
\end{equation}
\begin{equation}
u_{t}+uu_{x}+vu_{y}-H^{x}H_{x}^{x}-H^{y}H_{y}^{x}+gh_{x}=b_{x},
\end{equation}
\begin{equation}
v_{t}+uv_{x}+vv_{y}-H^{x}H_{x}^{y}-H^{y}H_{y}^{y}+gh_{y}=b_{y},
\end{equation}
\begin{equation}
H_{t}^{x}+uH_{x}^{x}+vH_{y}^{x}-u_{x}H^{x}-u_{y}H^{y}=0,
\end{equation}
\begin{equation}
H_{t}^{y}+uH_{x}^{y}+vH_{y}^{y}-v_{x}H^{x}-v_{y}H^{y}=0,
\end{equation}
\begin{equation}
h_{x}H^{x}+hH_{x}^{x}+h_{y}H^{y}+hH_{y}^{y}=0.
\end{equation}
\end{subequations}
Assuming that all dependent functions only depend on the single space variable~$x$,
the latter system brought to the form
\begin{subequations}
\label{eq:jul30.1}
\begin{equation}
h_{t}+uh_{x}+hu_{x}=0,\label{Eul_1d_cont}
\end{equation}
\begin{equation}
u_{t}+uu_{x}-H^{x}H_{x}^{x}+gh_{x}=b^{\prime},
\end{equation}
\begin{equation}
v_{t}+uv_{x}-H^{x}H_{x}^{y}=0,\label{eq:aug01.2}
\end{equation}
\begin{equation}
H_{t}^{x}+uH_{x}^{x}-u_{x}H^{x}=0,\label{Eul_1d_Hx}
\end{equation}
\begin{equation}
H_{t}^{y}+uH_{x}^{y}-v_{x}H^{x}=0,\label{eq:aug01.3}
\end{equation}
\begin{equation}
h_{x}H^{x}+hH_{x}^{x}=0.\label{Eul_divH}
\end{equation}
\end{subequations}
 Notice that by means of~(\ref{Eul_1d_cont}) and~(\ref{Eul_divH})
equation~(\ref{Eul_1d_Hx}) can be rewritten as
\begin{equation}
h_{t}H^{x}+hH_{t}^{x}=(hH^{x})_{t}=0.\label{Eul_1d_Hx_short}
\end{equation}
Hence, due to (\ref{Eul_divH}) one derives that
\begin{equation}
hH^{x}=a,\label{hHx_is_const}
\end{equation}
where $a$ is constant.

\section{The one-dimensional SMHD equations in Lagrangian coordinates}
\label{sec:02}

Similar to the gas dynamics equations \cite{bk:Ovsyannikov[2003]}
one can introduce mass Lagrangian coordinates $(s,t)$, where $x=\varphi(s,t)$,
and
\begin{equation}
\varphi_{t}(s,t)=\tilde{u}(s,t),\qquad\varphi_{s}(s,t)=\frac{1}{\tilde{h}(s,t)}.\label{eq:jul30.2}
\end{equation}
Here $\tilde{u}(s,t)=u(\varphi(s,t),t)$, $\tilde{h}(s,t)=h(\varphi(s,t),t)$.
The sign tilde~`$\ \tilde{}\ $' is omitted in further formulas. System
of equations (\ref{eq:jul30.1}) in mass Lagrangian coordinates
reduces to the equations
\begin{subequations}
\label{lagrMassSysConserved}
\begin{equation}
\left(\frac{1}{h}\right)_{t}-u_{s}=0,\label{lagrMassSysConserved_a}
\end{equation}
\begin{equation}
u_{t}-a^{2}\left(\frac{1}{h}\right)_{s}-ghh_{s}=b^{\prime},
\label{lagrMassSysConserved_b}
\end{equation}
\begin{equation}
v_{t}-aH_{s}^{y}=0,
\end{equation}
\begin{equation}
H_{t}^{y}-av_{s}=0,
\end{equation}
\end{subequations}
where due to~(\ref{hHx_is_const}) equations~(\ref{Eul_1d_Hx}) and~(\ref{Eul_divH}) are excluded
as identities.

In system~(\ref{lagrMassSysConserved}), the bottom topography
is described by the function~$b(x)$, where~$x=\varphi(s,t)$, and
the following relation
for the differentials~$dt$, $ds$ and~$dx$ holds~\cite{bk:YanenkRojd[1968]}
\begin{equation}
ds=h\,dx-hu\,dt,
\end{equation}
that means
\begin{equation}
s_{t}=-hu,\qquad s_{x}=h.\label{StSx}
\end{equation}

The general solution of the last two equations of system~(\ref{lagrMassSysConserved})
follows from the~d'Alembert formula for the wave equation and has
the form
\begin{equation}
v=f_{1}(s+at)+f_{2}(s-at),\qquad H^{y}=f_{1}(s+at)-f_{2}(s-at),\label{eq:aug03.1}
\end{equation}
where $f_{1}$ and $f_{2}$ are arbitrary functions of their arguments.

In variables~$t$, $s$, $\varphi$ system~(\ref{lagrMassSysConserved}) reduces to the only second-order~PDE:
\begin{equation}
\varphi_{tt}-\left(a^{2}\varphi_{s}-\frac{g}{2\varphi_{s}^{2}}\right)_{s}=b^{\prime}.\label{LagrEqns_Phi}
\end{equation}
Therefore, the study of equations~(\ref{eq:jul30.1}) in mass Lagrangian
coordinates is reduced to the analysis of a single equation~(\ref{LagrEqns_Phi}).

Notice that if $a=0$, then equation~(\ref{LagrEqns_Phi}) corresponds
to the classical one-dimensional shallow water equation, considered
in mass Lagrangian coordinates. Symmetries and conservation laws of
this case~($a=0$) have already been studied in~\cite{bk:KaptsovMeleshko_1D_classf[2018]}~(see
also~\cite{bk:SiriwatKaewmaneeMeleshko2016,bk:AksenovDruzkov_classif[2019]}).
Hence, for further analysis it is assumed that~$a\neq0$.

\begin{remark} \label{rem:btm_transform} In mass Lagrangian coordinates,
equations for inclined bottom~$b^{\prime}=b_{0}=\text{const}$ reduce
to equations for a horizontal bottom by means of the following transformation~\cite{bk:SWMHD_Karelsky2014,dorodnitsyn2019shallow}
\begin{equation}
\varphi=\tilde{\varphi}+\frac{b_{0}t^{2}}{2}.
\end{equation}
\end{remark}

\section{Lie Group classification}
\label{sec:03}

For the sake of simplicity in the present section we use the notation
$c=b^{\prime}$.

Calculations yield the following equivalence group:
\[
X_{1}^{e}=\frac{\partial}{\partial s},\,\,\,X_{2}^{e}=\frac{\partial}{\partial t},\,\,\,X_{3}^{e}=\frac{\partial}{\partial\varphi},
\]
\[
X_{4}^{e}=2\varphi\frac{\partial}{\partial\varphi}+s\frac{\partial}{\partial s}+t\frac{\partial}{\partial t}+3g\frac{\partial}{\partial g},\,\,\,X_{5}^{e}=2\varphi\frac{\partial}{\partial\varphi}+4s\frac{\partial}{\partial s}+t\frac{\partial}{\partial t}+3a\frac{\partial}{\partial a},
\]
\[
X_{6}^{e}=\varphi\frac{\partial}{\partial\varphi}+s\frac{\partial}{\partial s}+t\frac{\partial}{\partial t}-c\frac{\partial}{\partial c}.
\]
We also use the following involutions:
\[
(a):\,\,\,\varphi\mapsto-\varphi,\,\,\,g\mapsto-g,\,\,\,c\mapsto-c,
\]
\[
(b):\,\,\,t\mapsto-t,\,\,\,a\mapsto-a.
\]

\begin{remark} A shift of the bottom function $b=\tilde{b}+k$, where
$k$ is constant, does not change equations (\ref{eq:jul30.1}). \end{remark}

The equivalence transformations allow reducing certain constants
in the results of symmetry classification. The group classification
of equation~(\ref{LagrEqns_Phi}) considered for the case~$ag\neq0$.

An admitted generator is sought in the form
\[
X=\xi^{t}\frac{\partial}{\partial t}+\xi^{s}\frac{\partial}{\partial s}+\zeta^{\varphi}\frac{\partial}{\partial\varphi},
\]
which is prolonged for derivatives by standard formulas of the group
analysis~\cite{bk:Ovsiannikov1978}. Solving the determining equations
\begin{equation}
X\left(\varphi_{tt}-\left(a^{2}\varphi_{s}-\frac{g}{2\varphi_{s}^{2}}\right)_{s}-b^{\prime}\right)\Bigg|_{(\ref{LagrEqns_Phi})}=0,
\label{LagrEqnsDet}
\end{equation}
one derives the classifying equation
\begin{equation}
\zeta^{\prime\prime}-\zeta c^{\prime}-2k_{1}(\varphi c^{\prime}+c)=0,\label{eq:jul20.2}
\end{equation}
where
\begin{equation}
\zeta^{\varphi}=2k_{1}\varphi+\zeta(t),\,\,\,\xi^{s}=2k_{1}s+k_{2},\,\,\,\xi^{t}=2k_{1}t+k_{3},\label{eq:jul20.3}
\end{equation}
and $k_{i},i=1,2,3$ are arbitrary constants.

For any bottom function $b(x)$ equation~(\ref{LagrEqns_Phi}) admits the generators
\begin{equation}
X_{1}=\frac{\partial}{\partial s},\qquad X_{2}=\frac{\partial}{\partial t}.\label{kern}
\end{equation}
The Lie algebra consisting of these generators is called the kernel of admitted~Lie algebras of equation~(\ref{LagrEqns_Phi}).

Notice that differentiating equation (\ref{eq:jul20.2}) with respect
to $t$, one finds that
\begin{equation}
\zeta^{\prime}c^{\prime\prime}=0.\label{eq:jul20.9}
\end{equation}

Further analysis depends on a choice of the function $c$. According
to equation (\ref{eq:jul20.9}), one needs to consider $c^{\prime\prime}\neq0$
and $c^{\prime\prime}=0$. Results of this analysis are presented
in Table~\ref{tab:1}, where the first column lists the function~$c$
(up to equivalence transformations), the second column
gives the corresponding bottom topography, extensions of the kernel
of admitted Lie algebras are given in the third column. The generators
presented in Table~\ref{tab:1} are
\[
X_{1}=\frac{\partial}{\partial s},\,\,\,X_{2}=\frac{\partial}{\partial t},\,\,\,X_{3}=\frac{\partial}{\partial\varphi},\,\,\,X_{4}=t\frac{\partial}{\partial\varphi},
\]
\[
X_{5}=(\varphi+q)\frac{\partial}{\partial\varphi}+s\frac{\partial}{\partial s}+t\frac{\partial}{\partial t},\,\,\,X_{6}=\varphi\frac{\partial}{\partial\varphi}+s\frac{\partial}{\partial s}+t\frac{\partial}{\partial t},\,\,\,X_{7}=\left(\varphi+q_{1}\frac{t^{2}}{2}\right)\frac{\partial}{\partial\varphi}+s\frac{\partial}{\partial s}+t\frac{\partial}{\partial t},
\]

\[
X_{8}=e^{kt}\frac{\partial}{\partial\varphi},\,\,\,X_{9}=e^{-kt}\frac{\partial}{\partial\varphi},\,\,\,X_{10}=\cos{kt}\,\frac{\partial}{\partial\varphi},\,\,\,X_{11}=\sin{kt}\,\frac{\partial}{\partial\varphi}.
\]

\begin{table}[ht]
\centering
\def\arraystretch{1.6}%
\begin{tabular}{|c|c|c|}
\hline
$c(\varphi)$  & $b(x)$  & Extension \\
\hline
\hline
$0$  & $q_{1}$  & $X_{3},X_{4},X_{6}$ \\
\hline
 $\varphi^{-1}$ & $\ln{x}$  & $X_{5}$ \\
\hline
$\varphi$  & $\frac{1}{2}x^{2}+qx$  & $X_{8},X_{9}$ \\
\hline
$-\varphi$  & $-\frac{1}{2}x^{2}+qx$  & $X_{10},X_{11}$ \\
\hline
\end{tabular}

\caption{Group classification}
\label{tab:1}
\end{table}

\section{Lagrangian and Hamiltonian formalism of equation (\ref{LagrEqns_Phi}) }
\label{sec:04}

\subsection{Lagrangian}

Equation (\ref{LagrEqns_Phi}) can be represented as the Euler--Lagrange
equation. For this purpose one has to find a Lagrangian of the form
\begin{equation}
\mathcal{L}=\mathcal{L}(t,s,\varphi, \varphi_t,{\varphi}_s)\label{lagrGenForm}
\end{equation}
such that the equation
\begin{equation}
\displaystyle \frac{\delta\mathcal{L}}{\delta\varphi} = 0,
\label{lagrvarPhi}
\end{equation}
is equivalent to equation~(\ref{LagrEqns_Phi}), where
\begin{equation}
\frac{\delta}{\delta\varphi} =
\frac{\partial}{\partial\varphi}
 - D_t\left(\frac{\partial}{\partial{\varphi}_t}\right)
 - D_s\left(\frac{\partial}{\partial{\varphi}_s}\right)
\end{equation}
is the variational derivative, and $D_t$, $D_s$ are the total derivatives with respect to~$t$ and~$s$.
This is called the Helmholtz problem~\cite{bk:Douglas1939}.

\medskip{}

Substituting~$\varphi_{tt}$ found from~(\ref{LagrEqns_Phi}) into~(\ref{lagrvarPhi}),
and splitting it with respect to parametric derivatives,
one derives an overdetermined system
of differential equations in which~$t$,~$s$, $\varphi$, $\varphi_t$ and~$\varphi_s$
are considered as independent variables. Solving the resulting system
of equations, one finally arrives at the Lagrangian
\begin{equation}
{\displaystyle \mathcal{L}=\frac{\varphi_{t}^{2}}{2}-\frac{a^{2}\varphi_{s}^{2}}{2}-\frac{g}{2\varphi_{s}}+b.}\label{lagr}
\end{equation}

One of applications of the Lagrangian consists of
deriving conservation laws using Noether's identity, which
explicitly shows that invariance of the Lagrangian on solutions of the Euler--Lagrange
equation yields conservation laws for this equation (for example,
see~\cite{bk:Ibragimov[1983]}):
\begin{equation} \label{ident}
  X{\cal L} + {\cal L} D_j\xi^j = (\eta - \xi^j \varphi_j)  \frac {\delta
  {\cal L}}{\delta \varphi_j}  + D_j \left(\xi^j {\cal L} + (\eta - \xi^j \varphi_j){
\partial {\cal L}\over
\partial \varphi_j } \right),
\end{equation}
where $j\in\{t,s\}$, and the invariance of~Lagrangian means that the left hand
side of the latter identity is divergent.

\subsection{Hamiltonian}

The Lagrangian (\ref{lagr}) can be written in the form
\begin{equation}
{\cal L}(x,x_{s},\dot{x})=\frac{1}{2}\dot{x}^{2}+F(x,x_{s}),
\label{eq:Lagrangian_gen}
\end{equation}
where ${\displaystyle F=-\left(\frac{a^{2}x_{s}^{2}}{2}+\frac{g}{2x_{s}}-b(x)\right)}$,
dot ` $\dot{}$ ' means the derivative with respect to~$t$. Introducing
the variable~$\xi=\dot{x}$, the Lagrangian~${\cal L}$ becomes
\[
\tilde{{\cal L}}(\xi,x,x_{s})=\frac{1}{2}\xi^{2}+F(x,x_{s}).
\]
The Euler--Lagrange equation
\begin{equation}
\ddot{x}=\frac{\delta F}{\delta x}\label{eq:Euler-Lagrange}
\end{equation}
 in the Euler--Lagrange form can be rewritten as
\begin{equation}
\dot{\eta}=\frac{\delta\tilde{{\cal L}}}{\delta x},
\qquad
\dot{x}=\xi,\label{eq:2}
\end{equation}
where $\xi$ is found from the equation
\[
\eta=\frac{\delta\tilde{{\cal L}}}{\delta\xi}=\xi.
\]

As the Lagrangian~(\ref{eq:Lagrangian_gen}) is nonsingular~\cite{bk:Mokhov},
then one can derive the Hamiltonian as follows~\cite{bk:Ostrogradsky}.

Using the Legendre transformation
\[
\mathcal{H}=\dot{x}{\cal L}_{\dot{x}}-{\cal L}=\frac{1}{2}\dot{x}^{2}-F,
\]
the Hamiltonian becomes
\begin{equation}
\begin{array}{c}
\mathcal{H}(\eta,x,x_{s})=\frac{1}{2}\eta^{2}-F(x,x_{s}).\end{array}\label{eq:3}
\end{equation}
The Hamiltonian equations are
\begin{equation}
\dot{x}=\frac{\delta \mathcal{H}}{\delta\eta},
\qquad
\dot{\eta}=-\frac{\delta \mathcal{H}}{\delta x}.\label{eq:Hamiltonian}
\end{equation}
Substituting the Hamiltonian~(\ref{eq:3}) into~(\ref{eq:Hamiltonian}),
they are
\[
\dot{x}=\eta,
\qquad
\dot{\eta}=\frac{\delta F}{\delta x}.
\]
Hence, one notes that Hamiltonian equations~(\ref{eq:Hamiltonian})
coincide with the Euler--Lagrange equations~(\ref{eq:Euler-Lagrange}).

Recall that the canonical Hamiltonian equations~(we denote $x=q, \eta=p$)
\begin{equation}
\dot{q}  = { \partial \mathcal{H} \over  \partial {p} } , \qquad
\dot{p} = - { \partial \mathcal{H} \over  \partial {q}  },
\end{equation}
can be obtained by varying the action functional
\begin{equation}
\delta   \int_{t_1} ^{t_2} \left(
 p   \dot{q}   - \mathcal{H} ( t, { q} , { p} )
\right) dt = 0
\end{equation}
in the phase space $( { q}, { p} )$ (see, for
example,~\cite{bk:Ostrogradsky,bk:GelfandFomin_CalculusOfVariations}).

Lie point symmetries have to be written in the variables~$(t,{ q},{ p})$ as
\begin{equation}
X = \xi ( t, { q}, { p} )  { \partial \over \partial t } + \eta  (
t, { q},  { p} )  { \partial \over \partial q } + \zeta  ( t, { q},
{ p})  { \partial \over \partial p }.
\end{equation}

The newly established Hamilton identity derived in~\cite{bk:Dorod_Hamilt[2011]}
\begin{equation*}
 {\displaystyle
   \zeta   \dot{q}   + p   D_t \eta
-  X \mathcal{H} - \mathcal{H}  D_t\xi
\equiv   \xi \left( D_t\mathcal{H} - { \partial \mathcal{H} \over  \partial t } \right)  } -   \eta
    \left(  \dot{p}  +  { \partial \mathcal{H} \over  \partial q   } \right)
+  \zeta
    \left(  \dot{q}   - { \partial \mathcal{H} \over  \partial p } \right)
+ D_t \left[    p   \eta   - \xi \mathcal{H}  \right]
\end{equation*}
shows the relation between invariance of the~Hamiltonian~(which means
the left hand side of the identity is zero) and first integral on the
solution of Hamilton equations. Notice that the first expression in
the right hand side of the identity is zero on the solution of
Hamilton equations.

Further we use the Lagrangian approach to find conservation laws.

\subsection{Conservation laws}

In the present section, the conservation laws for equation~(\ref{LagrEqns_Phi})
in mass~Lagrangian coordinates and equations~(\ref{lagrMassSysConserved_a}),~(\ref{lagrMassSysConserved_b})
in~Eulerian coordinates, found by applying~Noether's theorem, are presented.

\smallskip{}

The local conservation law of a studied system of equations has the form of
a divergent expression that vanishes on solutions of the system,
\begin{equation}
(T^{1})_{t}+(T^{2})_{s}=0,
\end{equation}
where the conserved quantities $T^{1}$ and $T^{2}$ are usually called
the density and the flux of the conservation law.

\medskip{}

As in the previous section, here~$c=b^{\prime}$. Each conservation
law is preceded by the symmetry which was used in applying ~Noether's
theorem. The case~$c=q_{1}$ is not considered further, because of~Remark~\ref{rem:btm_transform},
it reduces to the case~$c=0$.
\begin{itemize}
\item Case $c$ is arbitrary.
\begin{equation}
X_{1}:\quad\left(\varphi_{t}\varphi_{s}
\right)_{t}-\left(\frac{\varphi_{t}^{2}+a^{2}\varphi_{s}^{2}}{2}-\frac{g}{\varphi_{s}}+b
\right)_{s}=0,\label{CLlagrX1}
\end{equation}
\begin{equation}
X_{2}:\quad\left(\frac{\varphi_{t}^{2}+\varphi_{s}^{2}}{2}+\frac{g}{2\varphi_{s}}
-b\right)_{t}+\left(
\frac{g\varphi_{t}}{2\varphi_{s}^{2}}-\varphi_{t}\varphi_{s}
\right)_{s}=0.\label{CLlagrX2}
\end{equation}
\item Case $c=0$.
\begin{equation}
X_{3}:\quad\left(\varphi_{t}\right)_{t}+\left(\frac{g}{2\varphi_{s}^{2}}-a^{2}\varphi_{s}\right)_{s}=0,
\end{equation}
\begin{equation}
X_{4}:\quad\left(t\varphi_{t}-\varphi\right)_{t}+\left(\frac{tg}{2\varphi_{s}^{2}}-ta^{2}\varphi_{s}\right)_{s}=0.\label{CLlagrX4}
\end{equation}
\item Case $c=\varphi$.
\begin{equation}
X_{8}:\quad\left(\left(\varphi-\varphi_{t}-\varphi_{s}\right)e^{t}\right)_{t}+\left(\left(a^{2}\varphi_{s}+\varphi_{t}+\varphi-\frac{g}{2\varphi_{s}^{2}}\right)e^{t}\right)_{s}=0,\label{CLlagrX8}
\end{equation}
\begin{equation}
X_{9}:\quad\left(\left(\varphi_{t}+\varphi_{s}+\varphi\right)e^{-t}\right)_{t}+\left(\left(\varphi-a^{2}\varphi_{s}-\varphi_{t}+\frac{g}{2\varphi_{s}^{2}}\right)e^{-t}\right)_{s}=0.\label{CLlagrX9}
\end{equation}
\item Case $c=-\varphi$.
\begin{equation}
X_{10}:\quad\left(\varphi\sin{t}+\varphi_{t}\cos{t}\right)_{t}-\left(\left(a^{2}\varphi_{s}-\frac{g}{2\varphi_{s}^{2}}\right)\cos{t}\right)_{s}=0,\label{CLlagrX10}
\end{equation}
\begin{equation}
X_{11}:\quad\left(\varphi\cos{t}-\varphi_{t}\sin{t}\right)_{t}+\left(\left(a^{2}\varphi_{s}-\frac{g}{2\varphi_{s}^{2}}\right)\sin{t}\right)_{s}=0.\label{CLlagrX11}
\end{equation}
\end{itemize}
\medskip{}

Conserved quantities $(^{l}T^{1},{}^{l}T^{2})$ and $(^{e}T^{1},{}^{e}T^{2})$
in Lagrangian and Eulerian coordinates are related as
\begin{equation}
^{e}T^{1}=h\,{}^{l}T^{1},\qquad{}^{e}T^{2}=hu\,{}^{l}T^{1}+{}^{l}T^{2}.
\end{equation}
Thus, the following conservation laws for equations~(\ref{lagrMassSysConserved_a})
and~(\ref{lagrMassSysConserved_b}) in~Eulerian coordinates are
obtained. The operators of total differentiation with respect to~Eulerian
coordinates~$t$ and~$x$ are denoted as~$D_{t}$ and~$D_{x}$.
\begin{itemize}
\item Case $c$ is arbitrary.
\begin{equation}
X_{1}:\quad D_{t}\left(2u\right)+D_{x}\left(u^{2}-(H^{x})^{2}+2gh-2b\right)=0,
\end{equation}
\begin{equation}
X_{2}:\quad D_{t}\left\{ (u^{2}+(H^{x})^{2}-2b)h+gh^{2}\right\} +D_{x}\left\{ \left(u^{2}-(H^{x})^{2}+2gh-2b\right)hu\right\} =0.
\end{equation}
\item Case $c=0$.
\begin{equation}
X_{3}:\quad D_{t}\left(uh\right)+D_{x}\left((u^{2}-(H^{x})^{2})h+\frac{gh^{2}}{2}\right)=0,
\end{equation}
\begin{equation}
X_{4}:\quad D_{t}\left(\left(tu-x\right)h\right)+D_{x}\left(\left(t(u^{2}-(H^{x})^{2})-xu\right)h+\frac{tgh^{2}}{2}\right)=0.
\end{equation}
\item Case $c=x$.
\begin{equation}
X_{8}:\quad D_{t}\left\{ \left(\left(x-u\right)h-1\right)e^{t}\right\} +D_{x}\left\{ \left(x+\left((H^{x})^{2}-u^{2}+xu\right)h-\frac{gh^{2}}{2}\right)e^{t}\right\} =0,
\end{equation}
\begin{equation}
X_{9}:\quad D_{t}\left\{ \left(\left(x+u\right)h+1\right)e^{-t}\right\} +D_{x}\left\{ \left(x+\left(u^{2}-(H^{x})^{2}+xu\right)h+\frac{gh^{2}}{2}\right)e^{-t}\right\} =0.
\end{equation}
\item Case $c=-x$.
\begin{equation}
X_{10}:\quad D_{t}\left(h(u\cos{t}+x\sin{t})\right)+D_{x}\left\{ \left(u^{2}-(H^{x})^{2}+\frac{gh}{2}\right)h\cos{t}+xhu\sin{t}\right\} =0,
\end{equation}
\begin{equation}
X_{11}:\quad D_{t}\left(h(u\sin{t}-x\cos{t})\right)+D_{x}\left\{\left(u^{2}-(H^{x})^{2}+\frac{gh}{2}\right)h\sin{t}-xhu\cos{t}\right\}=0.
\end{equation}
\end{itemize}
\begin{remark} The transition from conservation
laws in coordinates~$t,s,\varphi$ to conservation laws in coordinates~$t,s,h,u$
for system~(\ref{lagrMassSysConserved}) can be carried out straightforward
using formulas~(\ref{eq:jul30.2}) and therefore is not given here.
The only difficulty is that the conservation laws whose densities
or fluxes explicitly include the variable~$x=\varphi$ should be supplemented
by equations~(\ref{eq:jul30.2}). This is true for the conservation
laws~(\ref{CLlagrX4})--(\ref{CLlagrX11}) and also for~(\ref{CLlagrX1})
and~(\ref{CLlagrX2}) in case $b^{\prime}\neq0$. For example, the
conservation law~(\ref{CLlagrX8}) in terms of the variables~$t,s,h,u$
should be written as
\begin{equation}
\left(\left(x-u-\frac{1}{h}\right)e^{t}\right)_{t}+\left(\left(x+u+\frac{a^{2}}{h}-\frac{gh^{2}}{2}\right)e^{t}\right)_{s}=0,\qquad x_{t}=u,\qquad x_{s}=\frac{1}{h}.
\end{equation}
\end{remark}

\section{Invariant and partially invariant solutions }
\label{sec:05}

The knowledge of admitted Lie group allows one to construct some invariant
and partially invariant solutions of equations (\ref{eq:jul30.1}).
The procedure for obtaining substantially different invariant solutions
is based on finding an optimal system of subalgebras of the admitted
Lie algebra of the studied equations~\cite{bk:Ovsiannikov[1993opt],bk:Ovsyannikov[1962]}.
Choosing a subalgebra, say $L$, one has to find universal invariant
of the subalgebra by solving the equation
\[
XJ=0,\ \forall X\in L,
\]
where $J$ depends on all dependent and independent variables.
Separating the universal invariant into two parts, one derives a representation
of the invariant solution, which, after substitution it into the original
equations, provides a reduced system of equations. The main advantage
of the reduced system is that it contains fewer independent variables.

The kernel of admitted Lie algebras~(\ref{kern}) gives two one-dimensional
subalgebras: $\{X_{1}\}$ and $\{X_{2}+\mathscr{D} X_{1}\}$. The subalgebra~$\{X_{1}\}$
does not provide invariant solutions, because invariant solutions in this case have the form~$\varphi=\varphi(t)$, which is impossible due to the requirement~$\varphi_{s}\neq0$ following
from~(\ref{eq:jul30.2}).
Solutions invariant with respect to the subalgebra~$\{X_{2}+\mathscr{D} X_{1}\}$ are called travelling wave
type solutions, which are discussed in the next section.

\subsection{Travelling wave type solutions of equation (\ref{LagrEqns_Phi})}
\label{sec:TravellingWaves}

Here the general case of arbitrary bottom topography is studied, while
the further sections consider specific invariant solutions based
on the group classification results.

A solution of a travelling wave type for equation (\ref{LagrEqns_Phi})
is defined by the assumption $\varphi=\varphi(z)$, where $z=s-\mathscr{D} t$.
Substituting the representation of the solution into equation (\ref{LagrEqns_Phi}),
it becomes
\[
\left(\mathscr{D}^{2}-a^{2}-\frac{g}{\varphi^{\prime\,3}}\right)\varphi^{\prime\prime}=b^{\prime}.
\]
Integrating it, due to the equivalence transformation corresponding
to the generator $\displaystyle\frac{\partial}{\partial b}$, one gets
\[
(\mathscr{D}^{2}-a^{2})(\varphi^{\prime})^{3}-2b\varphi^{\prime}+2g=0.
\]
The latter equation can be represented as~$\varphi^{\prime}=F(\varphi)$. As
$u=-\mathscr{D}\varphi^{\prime}$ and $h=1/\varphi^{\prime}$, then in Eulerian
coordinates $u(x,t)=-\mathscr{D} F(x)$ and $h(x,t)=F^{-1}(x)$, where the function
$F(x)$ satisfies the equation
\[
(\mathscr{D}^{2}-a^{2})F^{3}-2bF+2g=0.
\]
Thus, the travelling wave solution corresponds to a stationary solution
for the functions~$u$ and~$h$ in Eulerian coordinates. Let $\chi(x)$
be such that $\chi^{\prime}(x)=h(x)$. Substituting $s=\mathscr{D} t+\chi(x)$
into~(\ref{eq:aug03.1}), one also obtains the functions~$v(x,t)$ and~$H^{y}(x,t)$
in an explicit form in~Eulerian coordinates.

\subsection{Invariant solutions for extensions of the kernel}

Because the case $c=0$ corresponds to the gas dynamics equations in Lagrangian coordinates~\cite{bk:AndrKapPukhRod[1998]}, our study in the present section is restricted by the cases of logarithmic and parabolic bottom topographies.
In the case of two independent variables
consideration of subalgebras of dimensions two and higher leads to non-trivial exact solutions. According
to~Table~\ref{tab:1}, for logarithmic and parabolic bottom, equation~(\ref{LagrEqns_Phi}) admits three- and four-dimensional~Lie algebras. For all
such algebras, optimal systems of subalgebras are already known and
are given in~\cite{bk:PateraWinternitz1977}.

\begin{table}[ht]
\def\arraystretch{2.0}%
\centering %
\begin{tabular}{|c|c|l|}
\hline
$c(\varphi)$  & Subalgebra  & Invariant solution or reduction \\
\hline
\hline
\multirow{2}{*}{$\varphi^{-1}$} & $X_{6}$  & ${\displaystyle \varphi=tQ\left(z\right),\;z=\frac{s}{t},\quad\left(z^{2}-a^{2}-\frac{g}{{Q^{\prime}}^{3}}\right)Q^{\prime\prime}-\frac{1}{Q}=0}$ \\
\cline{2-3} \cline{3-3}
 & $X_{2}+\mathscr{D} X_{1}$  & ${\displaystyle \varphi=Q(s-\mathscr{D} t),\quad\left(\mathscr{D}^{2}-a^{2}\right){Q^{\prime}}^{2}+\frac{2g}{Q^{\prime}}-2\ln Q=\text{const}}$ \\
\hline
\multirow{2}{*}{$\varphi$} & $X_{2}+\mathscr{D} X_{1}$  & ${\displaystyle \varphi=Q(s-\mathscr{D} t),\quad\left(\mathscr{D}^{2}-a^{2}\right){Q^{\prime}}^{2}+\frac{2g}{Q^{\prime}}-Q^{2}=\text{const}}$ \\
\cline{2-3} \cline{3-3}
 & $X_{8}+\beta X_{9}+\mu X_{1}$  & ${\displaystyle \varphi=\left(C_{1}+\frac{s}{\mu}\right)e^{t}+\left(C_{2}+\frac{\beta s}{\mu}\right)e^{-t}}$ \\
\hline
\multirow{3}{*}{$-\varphi$} & $X_{2}+\mathscr{D} X_{1}$  & ${\displaystyle \varphi=Q(s-\mathscr{D} t),\quad\left(\mathscr{D}^{2}-a^{2}\right){Q^{\prime}}^{2}+\frac{2g}{Q^{\prime}}+Q^{2}=\text{const}}$ \\
\cline{2-3} \cline{3-3}
 & $X_{10}+\mu X_{1}$  & ${\displaystyle \varphi=C_{1}\sin t+\left(\frac{s}{\mu}+C_{2}\right)\cos t}$ \\
\cline{2-3} \cline{3-3}
 & $X_{11}+\mu X_{1}$  & ${\displaystyle \varphi=C_{1}\cos t+\left(\frac{s}{\mu}+C_{2}\right)\sin t}$ \\
\cline{2-3}
\hline
\end{tabular}\caption{Invariant solutions in mass Lagrangian coordinates}
\label{tab:invsols}
\end{table}

\subsubsection{Case $c=\varphi^{-1}$}

\smallskip

Consider the Lie algebra $\{X_{1},X_{2},X_{6}\}$. Its optimal system
of one-dimensional subalgebras consists of
\begin{equation}
\{X_{1}\},\;\{X_{6}\},\;\{X_{2}+\mathscr{D} X_{1}\}.
\end{equation}

\begin{itemize}
\item Consideration of the one-dimensional subalgebra~$\{X_{6}\}$ leads
to the solutions of the form
\begin{equation}
\varphi=tQ(z),\qquad z=s/t.\label{scalingQ}
\end{equation}
Substituting into~(\ref{LagrEqns_Phi}), one gets the ODE
\begin{equation}
\left(z^{2}-a^{2}-\frac{g}{{Q^{\prime}}^{3}}\right)Q^{\prime\prime}-\frac{1}{Q}=0.
\end{equation}
\item
As studied earlier in Section~\ref{sec:TravellingWaves}, the
subalgebra~$\{X_{2}+\mathscr{D} X_{1}\}$ leads to the solutions of the form
\begin{equation}
\varphi=Q(s-\mathscr{D} t).
\end{equation}
Substituting into~(\ref{LagrEqns_Phi}), one gets the reduction
\begin{equation}
\left(\mathscr{D}^{2}-a^{2}-\frac{g}{{Q^{\prime}}^{3}}\right)Q^{\prime\prime}-\frac{1}{Q}=0.
\end{equation}
Multiplying by $Q^{\prime}$ and integrating, one finds
\begin{equation}
\frac{1}{2}\left(\mathscr{D}^{2}-a^{2}\right){Q^{\prime}}^{2}+\frac{g}{Q^{\prime}}-\ln Q=C_{1},\label{reduct0}
\end{equation}
where $C_1$ and $C_2$ are arbitrary constants, which can be cancelled by equivalence transformations. Further in the present section~$C_{1}$ and~$C_{2}$ denote
constants of integration.

The following particular solution can be found in case~$\mathscr{D}=\pm a$, namely
\begin{equation}
\varphi=\frac{g(s\pm at+C_{2})}{W_{0}\left(g(s\pm at+C_{2})\exp(C_{1}g-1)\right)},\label{partsol01}
\end{equation}
where $W_{0}$ is the principal branch of the Lambert~$W$ function~\cite{bk:Corless1996}.
\end{itemize}

\subsubsection{Case $c=\varphi$}

\smallskip

The optimal system of one-dimensional subalgebras of the~Lie algebra~$\{X_{1},X_{2},X_{8},X_{9}\}$
is
\begin{equation}
\{X_{1}\},\;\{X_{2}+\mathscr{D} X_{1}\},\;\{X_{8}+\beta X_{9}+\mu X_{1}\}.
\end{equation}

\begin{itemize}

\item $\{X_{2}+\mathscr{D} X_{1}\}$ leads to the reduction
\begin{equation}
\left(\mathscr{D}^{2}-a^{2}-\frac{g}{{Q^{\prime}}^{3}}\right)Q^{\prime\prime}-Q=0,
\end{equation}
where $\varphi=Q(s-\mathscr{D} t)$. Multiplying by $Q^{\prime}$ and integrating,
one finds
\begin{equation}
\frac{1}{2}\left(\mathscr{D}^{2}-a^{2}\right){Q^{\prime}}^{2}+\frac{g}{Q^{\prime}}-\frac{Q^{2}}{2}=C_{1}.\label{reduct1}
\end{equation}

In case $\mathscr{D}=\pm a$ one gets the particular solution
\begin{equation}
\varphi=\zeta-\frac{2C_{1}}{\zeta},\qquad\zeta^{3}=3(s\pm at+C_{2})g+\sqrt{9(s\pm at+C_{2})^{2}g^{2}+8C_{1}^{3}}.\label{partsol02}
\end{equation}

\item $\{X_{8}+\beta X_{9}+\mu X_{1}\}$, $\mu\neq0$, leads to the reduction
\begin{equation}
Q^{\prime\prime}-Q=0,\qquad\varphi=\frac{s}{\mu}(e^{t}+\beta e^{-t})+Q(t).
\end{equation}
This gives the invariant solution
\begin{equation}
\varphi=\left(C_{1}+\frac{s}{\mu}\right)e^{t}+\left(C_{2}+\frac{\beta s}{\mu}\right)e^{-t}.\label{invSolLagr_Exp}
\end{equation}
\end{itemize}

\subsubsection{Case $c=-\varphi$}

\smallskip

The optimal system of one-dimensional subalgebras of the~Lie algebra~$\{X_{1},X_{2},X_{10},X_{11}\}$
is
\begin{equation}
\{X_{1}\},\;\{X_{2}+\mathscr{D} X_{1}\},\;\{X_{10}+\mu X_{1}\},\;\{X_{11}+\mu X_{1}\}.
\end{equation}

\begin{itemize}
\item For $\{X_{2}+\mathscr{D} X_{1}\}$, similar to~(\ref{reduct1}), one obtains
\begin{equation}
\frac{1}{2}\left(\mathscr{D}^{2}-a^{2}\right){Q^{\prime}}^{2}+\frac{g}{Q^{\prime}}+\frac{Q^{2}}{2}=C_{1}.
\end{equation}
In case $\mathscr{D}=\pm a$ there is the particular solution
\begin{equation}
\varphi=\zeta+\frac{2C_{1}}{\zeta},\qquad\zeta^{3}=-3(s\pm at+C_{2})g+\sqrt{9(s\pm at+C_{2})^{2}g^{2}-8C_{1}^{3}}.\label{partsol03}
\end{equation}
\item $\{X_{10}+\mu X_{1}\}$, $\mu\neq0$, leads to the reduction
\begin{equation}
Q^{\prime\prime}+Q=0,\qquad\varphi=\frac{s}{\mu}\,\cos t+Q(t).
\end{equation}
This gives the invariant solution
\begin{equation}
\varphi=C_{1}\sin t+\left(\frac{s}{\mu}+C_{2}\right)\cos t.\label{invSolLagr_trig1}
\end{equation}
\item Similar to the previous case, for the one-dimensional subalgebra $\{X_{11}+\mu X_{1}\}$
one derives the invariant solution
\begin{equation}
\varphi=C_{1}\cos t+\left(\frac{s}{\mu}+C_{2}\right)\sin t.\label{invSolLagr_trig2}
\end{equation}
\end{itemize}
The results are summarized in Table~\ref{tab:invsols}. The particular
solutions~(\ref{partsol01}), (\ref{partsol02}) and~(\ref{partsol03})
are not presented in the table because of their cumbersome forms.
As mentioned above, the subalgebra~$\{X_{1}\}$ gives no invariant
solutions. It is also assumed~$\mu \neq0$ throughout the table, as for~$\mu=0$ no invariant solutions. Notice that the traveling wave type
solutions corresponding to the subalgebra~$\{X_{2}+\mathscr{D} X_{1}\}$
are particular cases of the solutions studied in Section~\ref{sec:TravellingWaves}.

\subsubsection{Invariant solutions in Eulerian coordinates}

\smallskip

Invariant solutions obtained in Lagrangian coordinates can be rewritten in Eulerian coordinates. Among invariant solutions only the case of the scaling group corresponding to~$X_6$ requires special explanation. All other invariant solutions are given in~Table~\ref{tab:invsols2}.

\begin{table}[ht]
\def\arraystretch{1.8}%
 \centering %
\begin{tabular}{|c|c|c|}
\hline
$b(x)$  & Subalgebra  & Invariant solution or reduction \\
\hline
\hline
arbitrary  & $X_{2}+\mathscr{D} X_{1}$  & See Section~\ref{sec:TravellingWaves} \\
\hline
${\displaystyle \ln{x}}$  & $X_{6}$  & $\begin{array}{l}
{\displaystyle h=R^{\prime},\qquad u=z-\frac{R}{R^{\prime}},\qquad v=f_{1}(\zeta_{+})+f_{2}(\zeta_{-}),}\\
{\displaystyle H^{x}=\frac{a}{R^{\prime}},\qquad H^{y}=f_{1}(\zeta_{+})-f_{2}(\zeta_{-}),\qquad\zeta_{\pm}=\left(\frac{R}{{R^{\prime}}^{2}}\pm a\right)t,}\\
{\displaystyle \text{where}\quad g{R}^{\prime\prime}+\frac{(a^{2}-R^{2}){R}^{\prime\prime}}{{R^{\prime}}^{3}}-\frac{1}{z}=0,\quad R=R(z),\quad z=x/t.\vspace{2pt}}
\end{array}$ \\
\hline
${\displaystyle \frac{x^{2}}{2}}$  & $X_{8}+\beta X_{9}+\mu X_{1}$  & $\begin{array}{l}
{\displaystyle h=\frac{\mu}{e^{t}+\beta e^{-t}},\qquad u=\frac{1}{e^{2t}+\beta}\left((e^{2t}-\beta)x+2(C_{1}\beta-C_{2})e^{t}\right),}\\
{\displaystyle v=f_{1}(\zeta_{+})+f_{2}(\zeta_{-}),\quad H^{x}=\frac{a}{\mu}(e^{t}+\beta e^{-t}),\quad H^{y}=f_{1}(\zeta_{+})-f_{2}(\zeta_{-}),}\\
{\displaystyle \zeta_{\pm}=\left(\frac{xe^{t}}{e^{2t}+\beta}-\frac{C_{1}e^{2t}+C_{2}}{e^{2t}+\beta}\right)\mu\pm at.\vspace{2pt}}
\end{array}$\\
\hline
\multirow{2}{*}{${\displaystyle -\frac{x^{2}}{2}}$} & $X_{10}+\mu X_{1}$  & $\begin{array}{l}
{\displaystyle h=\frac{\mu}{\cos t},\qquad u=\frac{C_{1}}{\cos t}-x\tan t,\qquad v=f_{1}(\zeta_{+})+f_{2}(\zeta_{-}),}\\
{\displaystyle H^{x}=\frac{a\cos t}{\mu},\qquad H^{y}=f_{1}(\zeta_{+})-f_{2}(\zeta_{-}),}\\
{\displaystyle \zeta_{\pm}=\left(\frac{x}{\cos t}-C_{1}\tan t-C_{2}\right)\mu\pm at.\vspace{2pt}}
\end{array}$ \\
\cline{2-3} \cline{3-3}
 & $X_{11}+\mu X_{1}$  & $\begin{array}{l}
{\displaystyle h=\frac{\mu}{\sin t},\qquad u=x\cot t-\frac{C_{1}}{\sin t},\qquad v=f_{1}(\zeta_{+})+f_{2}(\zeta_{-}),}\\
{\displaystyle H^{x}=\frac{a\sin t}{\mu},\qquad H^{y}=f_{1}(\zeta_{+})-f_{2}(\zeta_{-}),}\\
{\displaystyle \zeta_{\pm}=\left(\frac{x}{\sin t}-C_{1}\cot t-C_{2}\right)\mu\pm at.\vspace{2pt}}
\end{array}$ \\
\cline{2-3}
\hline
\end{tabular}\caption{Some invariant solutions in Eulerian coordinates}
\label{tab:invsols2}
\end{table}

For representation of invariant solution corresponding to the logarithmic bottom
and the subalgebra~$\{X_{6}\}$ one solves~(\ref{scalingQ})
with respect to~$s$, namely,
\begin{equation} \label{s_eq_tR}
s=tR\left(z\right),
\end{equation}
where $z=x/t$ and $R=Q^{-1}$.

Differentiating~(\ref{s_eq_tR}) with respect to~$t$ and~$x$ and taking into account~(\ref{StSx}), one derives the following equations in~Eulerian coordinates
\begin{equation}
zR^{\prime}-R=hu,\qquad R^{\prime}=h.\label{RhuEqns}
\end{equation}
Then, one can write the function~$R$ explicitly as
\begin{equation}
R=\left(z-u\right)h.
\end{equation}
Substituting~(\ref{RhuEqns}) into~(\ref{eq:jul30.1}), one also
obtains the constraint
\begin{equation}
g{R}^{\prime\prime}+\frac{(a^{2}-R^{2}){R}^{\prime\prime}}{{R^{\prime}}^{3}}-\frac{1}{z}=0.\label{EulerScalingInvSolConstrt}
\end{equation}
The invariant solution is
\begin{equation}
\begin{array}{c}
{\displaystyle h=R^{\prime},\qquad u=z-\frac{R}{R^{\prime}},\qquad v=f_{1}(\zeta_{+})+f_{2}(\zeta_{-}),}\\
{\displaystyle H^{x}=\frac{a}{R^{\prime}},\qquad H^{y}=f_{1}(\zeta_{+})-f_{2}(\zeta_{-}),\qquad\zeta_{\pm}=\left(\frac{R}{{R^{\prime}}^{2}}\pm a\right)t,}
\end{array}
\end{equation}
where $z=x/t$, and the function~$R(z)$ must satisfy equation~(\ref{EulerScalingInvSolConstrt}).

\medskip{}

The remaining solutions in~Eulerian coordinates corresponding to~(\ref{invSolLagr_Exp}),
(\ref{invSolLagr_trig1}) and~(\ref{invSolLagr_trig2}) are derived
in a similar way.

\subsection{Simple waves}

According to the group analysis method, a simple wave is a partially
invariant solution, where all unknown functions depend on a single
function \cite{bk:Yanenko[1956],bk:Ovsiannikov1978,bk:Meleshko[2005]}. Riemann wave
is a prticular case of simple waves for equations which can be written
in Riemann invariants\footnote{Sometimes a simple wave is also called by a Riemann wave.}.
Simple waves of equations (\ref{eq:jul30.1}) are discussed here.

Consider equation (\ref{LagrEqns_Phi})
\[
\varphi_{tt}-\left(a^{2}\varphi_{s}-\frac{g}{2\varphi_{s}^{2}}\right)_{s}=b^{\prime}.
\]
In Eulerian coordinates this equation corresponds to the system of
equations
\begin{equation}
\begin{array}{c}
h_{t}+uh_{x}+hu_{x}=0,\\
\displaystyle
u_{t}+uu_{x}+\left(\frac{a^{2}}{h^{3}}+g\right)h_{x}=b^{\prime}.
\end{array}\label{eq:jul30.10}
\end{equation}
Equations (\ref{eq:jul30.10}) coincide with the gas dynamics equations
for isentropic flows, where the sound speed is
\[
\lambda=\sqrt{\frac{a^{2}}{h^{2}}+gh}.
\]

Let $\sigma(h)$ be a function such that $\displaystyle\sigma^{\prime}(h)=\frac{\lambda}{h}$.
Using the Riemann invariants \cite{bk:YanenkRojd[1968],bk:Ovsyannikov[2003]}\footnote{Riemann waves of equations (\ref{eq:jul30.10}) are discussed in \cite{bk:SWMHD_Karelsky2011,bk:SWMHD_Karelsky2014}.}
\[
r=u+\sigma(h),\,\,\,l=u-\sigma(h),
\]
equations (\ref{eq:jul30.10}) are rewritten in the form
\begin{equation}
r_{t}+(u+\lambda)r_{x}=b^{\prime},\,\,\,l_{t}+(u-\lambda)l_{x}=b^{\prime}.\label{eq:aug01.1}
\end{equation}

Consider case $b^{\prime}=0$. As in the classical gas dynamics 
for a Riemann wave one of the Riemann invariants is assumed to be
constant. For example, assume that $r=k$, where $k$ is constant.
In Lagrangian coordinates this Riemann wave is defined by the equation
\begin{equation}
\varphi_{t}+\sigma(1/\varphi_{s})=k.\label{eq:aug04.1}
\end{equation}
Solutions of this equation can be found by the Cauchy method (method
of characteristics). Comprehensive study of constructing solutions
of equations even more general than equation~(\ref{eq:aug04.1}) is
given in~\cite{bk:Smirnov1964}~(v.4). In particular, using notations
of~\cite{bk:Smirnov1964}, equation~(\ref{eq:aug04.1}) has the form
\[
f(p,q)=p+\sigma(1/q)-k=0,
\]
where ${\displaystyle \frac{d\sigma(1/q)}{dq}=-\sqrt{a^{2}+\frac{g}{q^{3}}}}$.
The complete integral is
\[
\varphi=s\alpha+t(k-\sigma(1/\alpha))+\beta,
\]
where $\alpha$ and $\beta$ are constant. Using the envelope, one
can construct singular and general integrals, as well as a solution
of a Cauchy problem. Thus, one obtains the general solution of
the~Riemann wave in Lagrangian coordinates.

\begin{remark} System of equations (\ref{eq:jul30.1}) only has
trivial solutions of simple wave type: it is reducible to an invariant
solution. One class of such invariant solutions is considered in \cite{bk:SWMHD_Karelsky2011,bk:SWMHD_Karelsky2014},
where the authors studied solutions of system (\ref{eq:jul30.1})
invariant with respect to the admitted generator $t\partial_{t}+x\partial_{x}$.

Indeed, for a nontrivial simple wave type solution of (\ref{eq:jul30.1})
one has to assume that $aH^{y}\neq const$. These solutions can be
represented in the form
\[
u=U(h),\,\,\,H^{x}=a/h,\,\,\,v=V(h),\,\,\,H^{y}=H(h),
\]
where $U(h)$, $V(h)$ and $H(h)$ are some functions. Substituting
these functions into equations~(\ref{eq:jul30.10}), one obtains
\begin{equation}
\begin{array}{c}
{\displaystyle h_{t}+Uh_{x}+hU^{\prime}h_{x}=0,\,\,\,U^{\prime}(h_{t}+Uh_{x})+\left(\frac{a^{2}}{h^{3}}+g\right)h_{x}=0,}\\
V^{\prime}(h_{t}+Uh_{x})-ah^{-1}H^{\prime}h_{x}=0,\,\,\,H^{\prime}(h_{t}+Uh_{x})-ah^{-1}V^{\prime}h_{x}=0.
\end{array}\label{eq:aug01.7}
\end{equation}
As $H^{y}\neq const$, then $h_{x}\neq0$. Eliminating $h_{t}$, found
from the first equation of (\ref{eq:aug01.7}), substituting it into
the other equations, and taking into account that $h_{x}\neq0$, one
derives
\begin{equation}
U^{\prime}\,^{2}=\frac{a^{2}}{h^{4}}+\frac{g}{h},\label{eq:aug01.8}
\end{equation}
\begin{equation}
V^{\prime}hU^{\prime}+ah^{-1}H^{\prime}=0,\,\,\,ah^{-1}V^{\prime}+H^{\prime}hU^{\prime}=0.\label{eq:aug01.9}
\end{equation}
As equations (\ref{eq:aug01.9}) compose a homogeneous system of linear
algebraic equations with respect to $H^{\prime}$ and $V^{\prime}$,
and due to $H^{\prime}\neq0$, one derives that
\[
U^{\prime}\,^{2}=\frac{a^{2}}{h^{4}}.
\]
Comparison of the latter relation with (\ref{eq:aug01.8}) gives the
contradiction to the condition that $g\neq0$.

\end{remark}

\section{Conclusion}

The present paper is devoted to the Lie group analysis of the
one-dimensional~SMHD equations within~Gilman's model. The~SMHD
equations are considered in cases of a plane and uneven bottom
topography.

We would like to outline three important results. Firstly, it is
shown that the system of equations written in Eulerian coordinates
reduces to a single second-order partial differential equation in
Lagrangian coordinates, while the rest of equations were integrated
explicitly. Complete Lie group classification with respect to bottom
topography of the single equation is performed.

Secondly, the~Lagrangian formalism and Noether's theorem are used to
construct conservation laws of the equations. For all cases of a
bottom topographies the conservation laws are constructed, including some
new conservation laws. These results are also represented in~Eulerian
coordinates.

Thirdly, invariant and partially invariant solutions are
constructed. The symmetry classification separated out all bottom
topographies into four classes. For each of the classes an
extension of the kernel of admitted Lie algebras is found and all
invariant solutions are considered. The kernel of admitted Lie
algebras consists of the shifts with respect to time and the mass
Lagrangian space coordinate that allows constructing traveling wave
solutions. It should be
noted that such solutions in Eulerian coordinates correspond to
stationary solutions for $h$, $u$ and $H^x$, whereas for $v$ and
$H^y$ we obtained the general solution, which is expressed through
$h$ in explicit form with two arbitrary functions.
Almost all invariant solutions are found in explicit
form both in mass Lagrangian coordinates and in Eulerian
coordinates. Analysis of~Riemann waves in mass Lagrangian
coordinates and comparison of them with  simple waves in~Eulerian
coordinates are presented in the paper.

\section*{Acknowledgements}

The research was supported by Russian Science Foundation Grant No.~18-11-00238
``Hydro\-dynamics-type equations: symmetries, conservation laws,
invariant difference schemes''. E.I.K. acknowledges Suranaree University
of Technology~(SUT) and Thailand Science Research and Innovation~(TSRI)
for Full-time Doctoral Researcher Fellowship~(Full-time61/15/2021).


\end{document}